\documentclass{article}

\usepackage[nonatbib]{neurips_data_2023}

\usepackage[utf8]{inputenc} 
\usepackage[T1]{fontenc}    
\usepackage{hyperref}       
\usepackage{url}            
\usepackage{booktabs}       
\usepackage{amsfonts}       
\usepackage{nicefrac}       
\usepackage{microtype}      
\usepackage{xcolor}         
\usepackage{bm}

\usepackage{enumitem}
\usepackage{multirow}
\usepackage{graphicx}
\usepackage{wrapfig}
\usepackage{color}
\usepackage{threeparttable}

\newcommand{\mylist}[1]{
\begin{itemize}[leftmargin=*] 
    \item {{#1}}
\end{itemize}}

\newcommand{\xh}{{\bm x}^{\rm h}}
\newcommand{\xp}[1]{{\bm x}^{\rm p}_{#1}}
\newcommand{\G}{{\mathcal G}}

\usepackage{makecell}

\title{LaDe: The First Comprehensive Last-mile Delivery Dataset from Industry}

 \author{%
  Lixia Wu$^{1, \dag}$, Haomin Wen$^{1,\dag}$, Haoyuan Hu$^{1}$, Xiaowei Mao$^{2,1}$, Yutong Xia$^{3}$, \\
  \textbf{Ergang Shan$^{1}$, Jianbin Zheng$^{1}$, Junhong Lou$^{1}$, Yuxuan Liang$^{4}$\thanks{$\dag$: Equal contribution. Y. Liang is the corresponding author of this paper. Email: yuxliang@outlook.com},} \\ 
  \textbf{Liuqing Yang$^{4}$, Roger Zimmermann$^{3}$, Youfang Lin$^{2}$, Huaiyu Wan$^{2}$} \\
  \textsuperscript{\rm 1}Cainiao Network  \textsuperscript{\rm 2}Beijing Jiaotong University \textsuperscript{\rm 3}National University of Singapore\\
   \textsuperscript{\rm 4}Hong Kong University of Science and Technology (Guangzhou)  \\
    \texttt{\{wallace.wulx,wenhaomin.whm,haoyuan.huhy,ergang.se\}@cainiao.com} \\
    \texttt{yuxliang@outlook.com;lqyang@ust.hk;yutong.xia@u.nus.edu}  \\
    \texttt{rogerz@comp.nus.edu.sg;\{maoxiaowei,hywan,yflin\}@bjtu.edu.cn}\\
}

\begin{document}  

\maketitle

\begin{abstract}
  Real-world last-mile delivery datasets are crucial for research in logistics, supply chain management, and spatio-temporal data mining. Despite a plethora of algorithms developed to date, no widely accepted, publicly available last-mile delivery dataset exists to support research in this field. In this paper, we introduce \texttt{LaDe}, the first publicly available last-mile delivery dataset with  millions of packages from the industry. \texttt{LaDe} has three unique characteristics:  (1) \textit{Large-scale}. It involves 10,677k packages of 21k couriers over 6 months of real-world operation. (2) \textit{Comprehensive information}. It offers original package information, such as its location and time requirements, as well as task-event information, which records when and where the courier is while events such as task-accept and task-finish events happen.  (3) \textit{Diversity}. The dataset includes data from various scenarios, including package pick-up and delivery, and from multiple cities, each with its unique spatio-temporal patterns due to their distinct characteristics such as populations. We verify \texttt{LaDe} on three tasks by running several classical baseline models per task. We believe that the large-scale, comprehensive, diverse feature of \texttt{LaDe} can offer unparalleled opportunities to researchers in the supply chain community, data mining community,  and beyond. The dataset homepage is publicly available at https://huggingface.co/datasets/Cainiao-AI/LaDe.
\end{abstract}


\vspace{-1em}
\section{Introduction} \label{sec:introduction}
\vspace{-1em}


\par Driven by increasing urbanization  and e-commerce development, last-mile delivery has emerged as a critical research area with growing interest from scholars and practitioners. \textbf{Last-Mile Delivery}, as illustrated in Figure~\ref{fig:data_illustration}, is the package transport process that connects the depot and the customers, including both the package pick-up \cite{macioszek2018first, ranathunga2021solution} and delivery \cite{boysen2021last, ratnagiri2022scalable} process. In addition to being a key to customer satisfaction, last-mile delivery is both the most expensive and time-consuming part of the shipping process \cite{olsson2019framework, mangiaracina2019innovative}. Consequently, researchers from different fields, from logistics operation management to spatio-temporal data mining, have been consistently shedding light on problems in last-mile delivery in recent years. These problems include route planning \cite{zeng2019last, li2021learning, almasan2022deep}, Estimated Time of Arrival (ETA) prediction \cite{wu2019deepeta, end2end2020Araujo, gao2021deep}, and route prediction \cite{e_le_me, wen2021package, wen2022graph2route}, etc. A quick search for ``last-mile delivery'' on Google Scholar returns over 19,400  papers since 2018.


\par Recent endeavors \cite{wu2019deepeta, end2end2020Araujo, gao2021deep} focus on leveraging machine/deep learning techniques for problems in last-mile delivery research.  A critical prerequisite for those researches is the availability of high-quality, large-scale datasets. Since such datasets have the potential to significantly accelerate advancements in specific fields,  such as ImageNet \cite{deng2009imagenet} for computer vision and GLUE \cite{wang2018glue} for natural language processing. Nonetheless, in the domain of last-mile background research, a multitude of algorithms have been devised, but there is still an absence of a widely recognized, publicly accessible dataset. Consequently, research in this field has become concentrated within a limited number of industrial research laboratories, thereby restricting transparency and hindering research progress. Moreover, the lack of public datasets also poses a hurdle for industry practitioners to develop advanced algorithms for last-mile delivery.



\par To meet the rising calling for a public dataset, we propose \texttt{LaDe}, the first comprehensive \underline{La}st-mile \underline{De}livery dataset collected by Cainiao. It contains both package pick-up and delivery data as depicted in Figure~\ref {fig:data_illustration}. \texttt{LaDe} has several merits: (1) \textit{Large-scale}, covering 10,677k packages of 21k couriers across 6 months. To the best of our knowledge, this is the largest publicly available dataset. (2) \textit{Comprehensive}, providing detailed information on package, location, task-event, and courier. (3) \textit{Diverse}, collecting data from both pick-up and delivery processes across various cities. By virtue of these advantages, \texttt{LaDe} can be employed to evaluate a wide spectrum of last-mile-related tasks. In this paper, we investigate its properties by three tasks, including route prediction \cite{e_le_me, wen2021package, wen2022graph2route}, estimated time of arrival prediction \cite{wu2019deepeta, end2end2020Araujo, gao2021deep}, and spatio-temporal graph forecasting \cite{li2018diffusion, yao2018deep, bai2020adaptive}. Beyond these tasks, it is easy to integrate some of the aforementioned features to support additional tasks. We believe that such a large-scale dataset like \texttt{LaDe} is a critical resource for developing advanced algorithms under the context of last-mile delivery, as well as for providing critical training and benchmarking data for learning-based algorithms. Overall, we identify three key contributions of this work:

\begin{figure}[t]
		\centering
		\includegraphics[width=1 \linewidth]{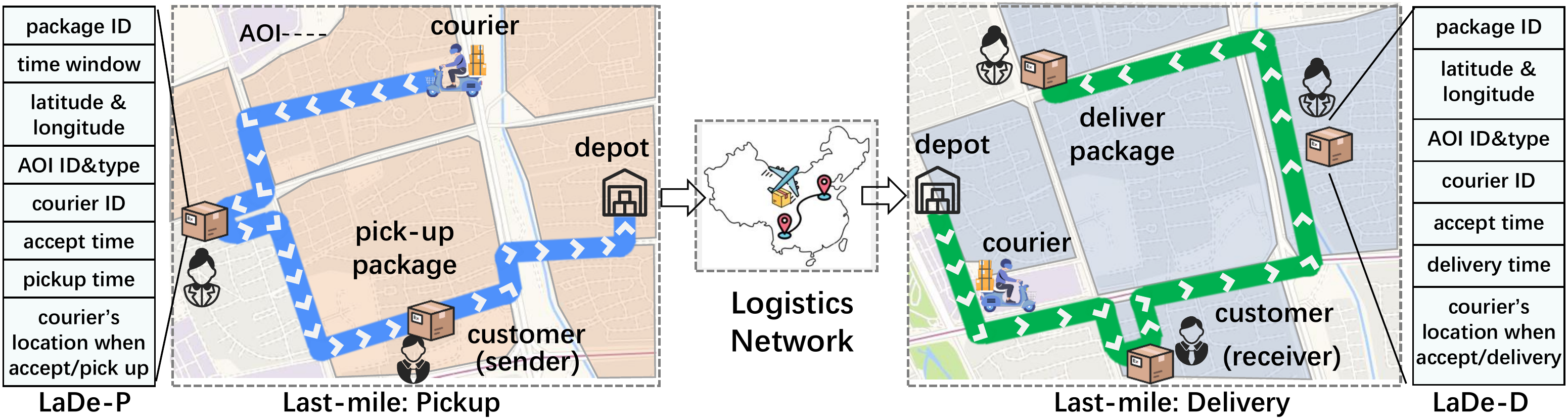}
		
		\caption{Overview of \texttt{LaDe} from last-mile delivery (better viewed in color), which includes two sub-datasets: \texttt{LaDe-P} from package pick-up process (i.e., couriers pick up packages from sender customers and return the depot) and \texttt{LaDe-D} from delivery process (i.e.,  couriers deliver packages from the depot to receiver customers). }
		\label{fig:data_illustration}
  \vspace{-1em}
\end{figure}

\begin{itemize}[leftmargin=*]
    \item \textbf{A New Dataset.} We collect, process, and release \texttt{LaDe}. The dataset boasts large-scale, comprehensive, and diverse characteristics. To the best of our knowledge, it is the first exhaustive, industry-scale last-mile delivery dataset. The dataset is publicly accessible at https://huggingface.co/datasets/Cainiao-AI/LaDe.
    \vspace{-0.5em}
    
    \item \textbf{Comprehensive Data Analysis.} Extensive data analysis is conducted to depict and highlight the properties of the dataset. Based on the analysis, we introduce potential tasks supported by \texttt{LaDe}, from logistics operation management to spatio-temporal data mining, and beyond.
    \vspace{-0.5em}

    \item \textbf{Benchmark on Real-World Tasks.} We benchmark this dataset by performing three representative tasks, including service route prediction, estimated time of arrival prediction, and spatio-temporal graph forecasting. The source codes for these tasks are provided to promote research in this field.
    \vspace{-0.5em}

\end{itemize}

\par The remainder of this paper is structured as follows. Section \ref{sec:related_work} discusses related work, and Section \ref{sec:dataset_intro} introduces the details of the dataset, including the methodology used to construct the dataset, and the statistics and properties of the dataset. In Section \ref{sec:application}, we benchmark the dataset on three tasks and discuss the potential use of the data in related research fields.

\section{Related Work} \label{sec:related_work}
\par \textbf{Dataset Perspective.} To the best of our knowledge, there is no publicly available last-mile dataset containing both package pick-up and delivery data. The most relative effort comes from Amazon \cite{merchan20222021} (named AmazonData in this paper). It is a courier-operated sequence dataset proposed for a last-mile routing research challenge hosted by Amazon. Specifically, this dataset contains 9,184 historical routes performed by Amazon couriers in 2018 in five metropolitan areas in the United States. Despite the contribution of AmazonData to the research field, it still has three limitations: 1) Without pick-up data, it only contains data generated in the package delivery process; 2) Small scale, in terms of spatio-temporal range and the number of trajectories; 3) Lack of courier-related and task-event-related information, which prevents it from benefiting a wider group of researchers with different interests. In light of the above issues, we introduce an industry-scale, comprehensive dataset (i.e., \texttt{LaDe}) for researchers to develop and evaluate new ideas on real-world instances in last-mile delivery.  The scale of \texttt{LaDe} is 5 times of AamazonData in terms of package number and 50 times in terms of trajectory number.  We provide a detailed comparison of AamazonData and \texttt{LaDe} in Table \ref{tab:data_relate_work}.

\begin{table}[htbp]
	\centering
	\caption{Comparison between \texttt{LaDe} and the related dataset.}
	\setlength\tabcolsep{2 pt}
	\resizebox{\linewidth}{!}
	{
		\begin{tabular}{c|ccccccccc}
			\toprule
			Dataset & Time span  & \#Trajectories & \#Couriers  & \#Packages & Delivery Data &  Pick-up Data & Courier Info & Task-event Info \\
			\midrule
			AmazonData &  4 months  & 9k & - & 2,182k & $\checkmark$ & $\times$ & $\times$ & $\times$ \\
			LaDe & 6 months   & {619k} & 21k  & 10,677k & $\checkmark$ & $\checkmark$ & $\checkmark$ & $\checkmark$ \\
			\bottomrule
		\end{tabular}
 		\label{tab:data_relate_work}
	}
	\label{datasets}
\end{table}


\par \textbf{Application Perspective.} Overall, last-mile logistics is an emerging interdisciplinary research area connecting transportation and AI technology, in which deep learning methods have long been the most popular model \cite{olsson2019framework}. Broadly speaking, there are four branches in this field:  1) Emerging trends and technologies, which focus on technological solutions and innovations in last-mile logistics, such as courier's route and arrival time prediction \cite{wen2022graph2route, gao2021deep}, self-service technologies \cite{vakulenko2018s}, drone-assisted delivery \cite{taniguchi2020modelling}. 2) Last-mile-related data mining \cite{ruan2022discovering, ruan2020learning}, which aims to excavate the underlying patterns of knowledge from data generated by real-world operations for better logistics management. 3) Operational optimization, which focuses on optimizing last-mile operations and making better operational decisions, such as vehicle routing problem \cite{zeng2019last, breunig2019electric}, delivery scheduling \cite{han2017appointment}, and facility location selection \cite{jahangiriesmaili2017solution, kedia2020locating}. 4) Supply chain structures, which focused on designing structures for last mile logistics, such as the network design \cite{lim2018examining}.  
We refer readers to the paper \cite{olsson2019framework} for a more detailed, systematic classification of last-mile-related research. The proposed \texttt{LaDe} contains instances based on real operational data that researchers can use to advance the state-of-the-art in their fields and to expand its applications to industry settings.


\section{Proposed Dataset: LaDe} \label{sec:dataset_intro}
\par In this section, we formally introduce the \texttt{LaDe} Dataset. First, we describe the data collection process, followed by a detailed discussion of \texttt{LaDe}'s data fields and dataset statistics. Finally, we conduct a comprehensive analysis to highlight its unique properties. The dataset can be freely downloaded at {https://huggingface.co/datasets/Cainiao-AI/LaDe} and noncommercially used with a custom license CC BY-NC 4.0\footnote{https://creativecommons.org/licenses/by-nc/4.0/}.

\subsection{Data Collection} \label{sec:data_collection}

\par This dataset is collected by Cainiao Network\footnote{https://global.cainiao.com/}, one of China's largest logistics platforms, which handles a tremendous volume of packages each day. A typical process for shipping a package involves the following steps: 1) The customer (sender) places a package pick-up order through the online platform. 2) The platform dispatches the order to an appropriate courier. 3) The courier picks up the package within the specified time window and returns to the depot (this constitutes the package pick-up process). 4) The package departs from the depot and traverses the logistics network until it reaches the target depot. 5) At the target depot, the delivery courier retrieves the package and delivers it to the recipient customer (known as the package delivery process). Among these steps, step 3 and 5 are referred to as the last-mile delivery, where couriers pick up/deliver packages from/to customers. Note that there is a notable difference between the pick-up and delivery scenarios. In the package delivery process, packages assigned to a particular courier are determined prior to the courier's departure from the depot. Conversely, in the pick-up process, packages assigned to a courier are not settled at the beginning. Rather, they are revealed over time, as customers can request pick-ups at any time. The dynamic nature of package pick-up presents substantial challenges in the research field.  To advocate more efforts for the challenge and make the data more diverse, \texttt{LaDe} contains two sub-datasets in both pick-up and delivery scenarios, named \texttt{LaDe-P} and \texttt{LaDe-D}, respectively.

\begin{wrapfigure}{R}
{0.3\textwidth}\vspace{-1em}
  \includegraphics[width= \linewidth]{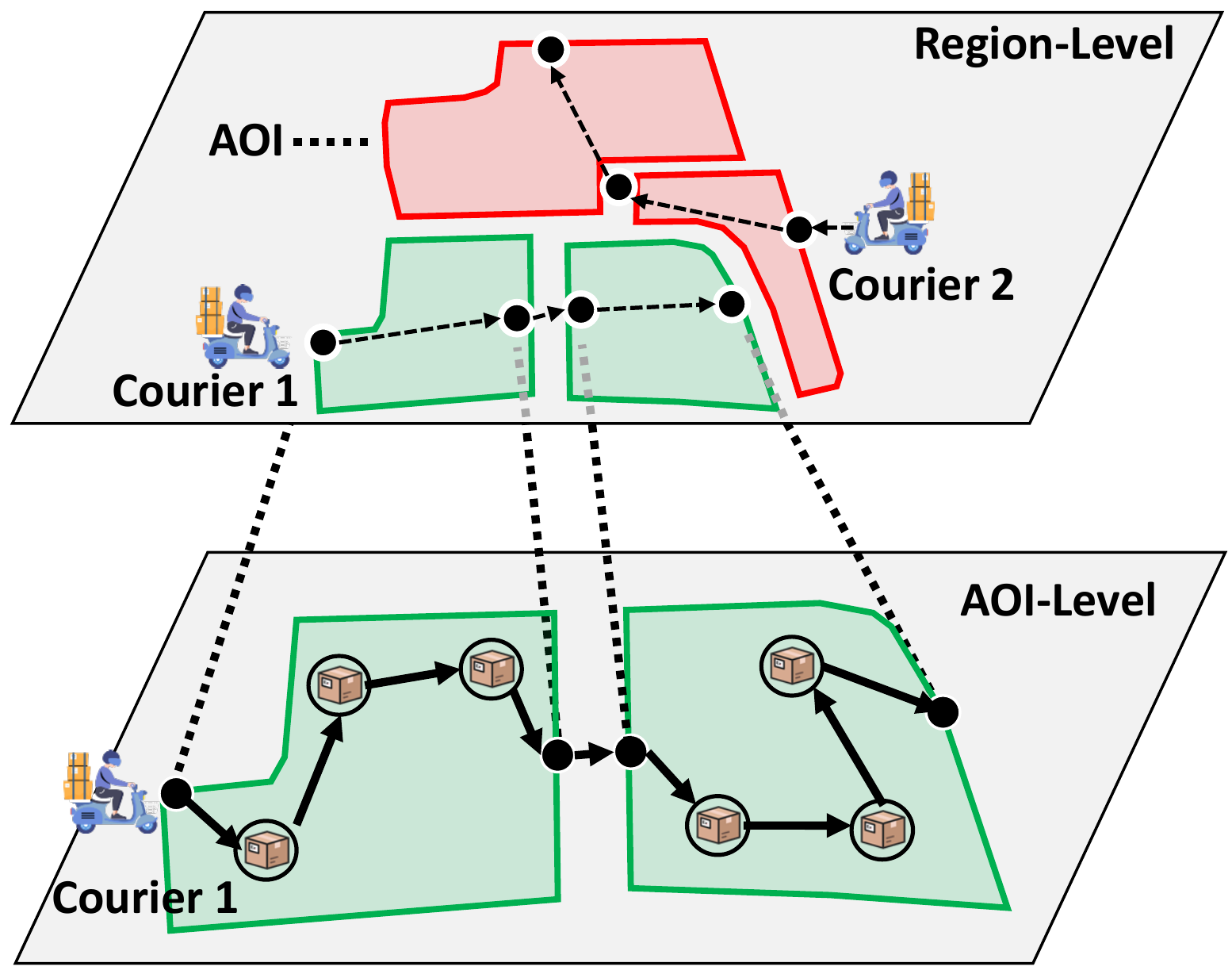}
  \vspace{-1.4em}
  \caption{Region-level and AOI-level data.}\label{fig:location_level}
  \vspace{-1em}
\end{wrapfigure}
\par Specifically, we collect millions of package pick-up/delivery data generated in 6 months from different cities in China. To increase the diversity, we carefully selected 5 cities - Shanghai, Hangzhou, Chongqing, Jilin, and Yantai - which possess distinct characteristics such as populations, more details can be found in Table~\ref{tab:city_information} of Appendix~\ref{appendix:data_statistics}. A city contains different regions, with each region  composed of several AOIs (Area of Interest) for logistics management. And a courier is responsible for picking up / delivering packages in several assigned AOIs. We give a simple illustration of the region-level and AOI-level segmentation of a city in Figure~\ref{fig:location_level}.  To collect the data for each city, we first randomly select 30 regions in the city. Subsequently, we randomly sample couriers in each region and pick out all the selected couriers' picked-up/delivery packages during the 6 months. To safeguard the privacy of both customers and couriers, we applied perturbations to the latitude and longitude points collected in the data.  The accuracy of the latitude and longitude is limited to 10 meters.  The data must not be assumed to indicate any of Cainiao's business interests. 

\subsection{Dataset Details \& Statistics}

\par In this subsection, we present the dataset details and its basic statistics. The detailed data field description of \texttt{LaDe-P} and \texttt{LaDe-D} can be found in Table \ref{tab:pickup_data_field} and Table \ref{tab:delivery_data_field} in Appendix~\ref{appendix:data_detail}.

\par To facilitate the utilization and analysis of the dataset, we transform and arrange each sub-dataset into tabular data presented in CSV format. Each record in this format contains relevant information pertaining to a picked-up or delivered package, primarily addressing the ``who, where, when'' aspects. Specifically, the record specifies which courier picked up or delivered the package, the location of the package, and the corresponding time. The recorded information can be broadly categorized into four types: 1) package information, which records the package ID and time windows requirements (if applicable); 2) stop information, recording the package's location information such as coordinates, AOI ID, and AOI type;  3) courier information, recording the courier's ID, and each courier is equipped with a personal digital assistant (PDA), which will consistently report the status of a courier (e.g., GPS) to the platform; 4) task-event information, recording the features of package accept, pick-up or delivery event, including when the event happens and the courier's location.


\par
\begin{figure}[!t]
	\centering
        \includegraphics[width=1\linewidth ]{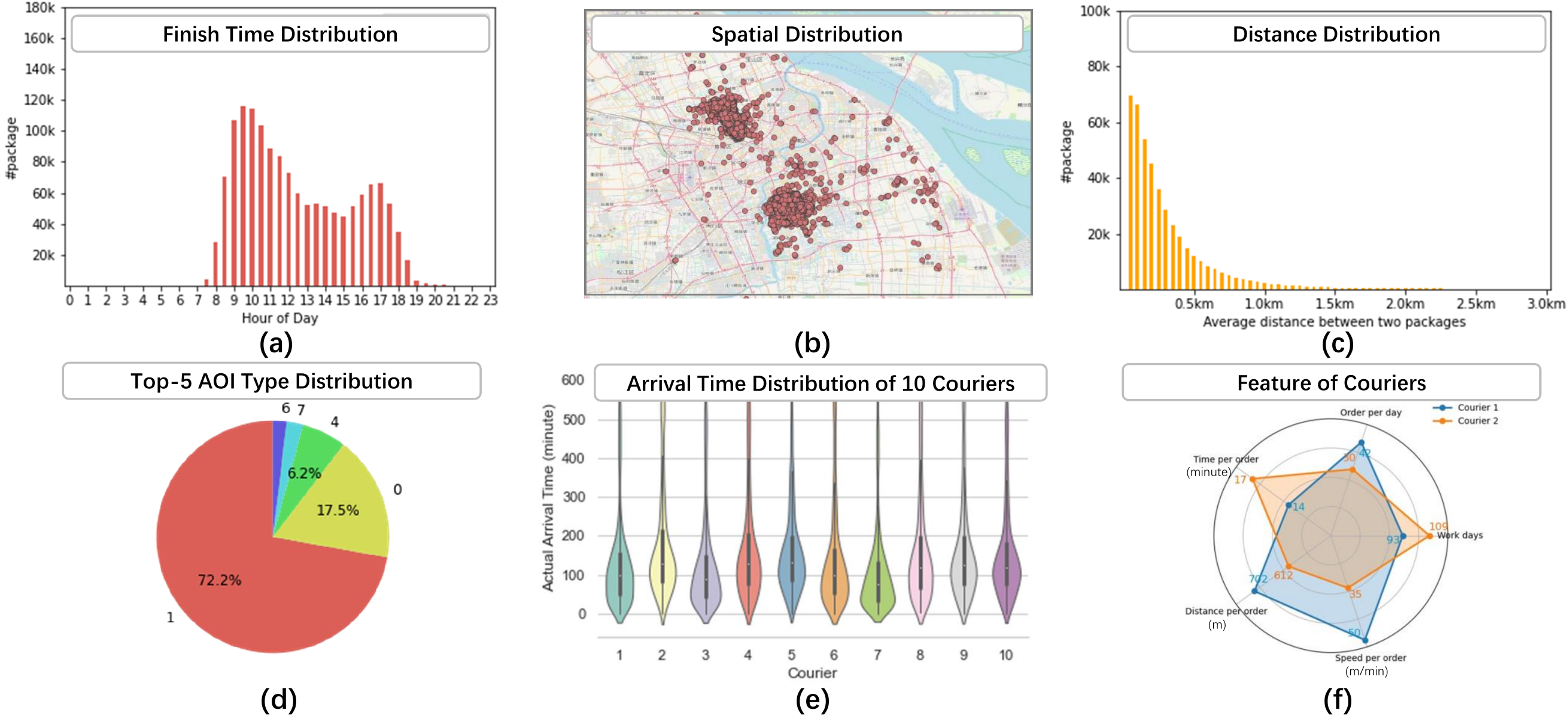}
	\caption{Spatial and temporal distribution of data in Shanghai of \texttt{LaDe-P}.}
	\label{fig:data_profile}
\end{figure}

 Overall, the package and task-event information can be recorded once the courier accepts the order, or finishes the order. Information about the stop comes from the geo-decoding system used in Cainiao, which can parse the input location address into its corresponding coordinates with a given accuracy. 
Table \ref{tab:pickup_data_statistics} shows the statistics of the \texttt{LaDe-P}. Due to the page limitation, please refer to Table~\ref{tab:delivery_data_statistics} in Appendix~\ref{appendix:data_statistics} for the  statistics of the \texttt{LaDe-D}.
Moreover, to intuitively illustrate the spatio-temporal characteristics of the dataset, we draw the spatial and temporal distribution of one city (Shanghai) in Figure~\ref{fig:data_profile} for one sub-dataset \texttt{LaDe-P}. From the Figure, we have the following observations. \textbf{Obs1:} Figure~\ref{fig:data_profile}(a) shows that couriers' work time starts from 8:00 and ends at 19:00. The volume of package pick-up has a peak at 9:00 am and 5:00 pm, respectively. \textbf{Obs2:} Figure~\ref{fig:data_profile}(b) and Figure~\ref{fig:data_profile}(c) shows the spatial distribution of packages, where the distance between consecutive packages in a courier's route is usually within 1km. \textbf{Obs3:} Figure~\ref{fig:data_profile}(d) shows the distribution of the top 5 AOI types in the data, illustrating that over 70\% packages come from type 1. \textbf{Obs4:} Figure~\ref{fig:data_profile}(e) shows the actual arrival time of 10 randomly selected couriers, from which we observed differences in the work efficiency of different couriers. It also shows that a majority of packages are picked up within 3 hours. \textbf{Obs5:} Figure~\ref{fig:data_profile}(f) depicts the profile of two couriers in the dataset, where different characteristics such as work days, and average orders per day are observed.


\begin{table}[!t]
	\centering
	\caption{Statistics of \texttt{LaDe-P}. AvgETA stands for the average arrival time per package. AvgPackage means the average package number of a courier per day. The unit of AvgETA is minute.}
	\setlength\tabcolsep{2 pt}
	\resizebox{\linewidth}{!}{
    	{
    		\begin{tabular}{cccccccccc}
    			\toprule
    			City & Time span & Spatial span & \#Trajectories & \#Couriers  & \#Packages &\#GPS points & AvgETA & AvgPackage\\
    			\midrule
    			Shanghai  &  6 months & 20km$\times$20km & 96k & 4,502  & 1,450k & 1,785k & 151 & 15.0 \\
    			Hangzhou  & 6 months  & 20km$\times$20km & 119k  & 5,347  & 2,130k & 2,427k & 146 & 17.8\\
    			Chongqing & 6 months  & 20km$\times$20km & 83k  & 2,982  & 1,172k & 1,475k & 140 & 14.0\\
    			Yantai    & 6 months  & 20km$\times$20km & 71k  & 2,593  & 1,146k & 1,641k & 137 & 16.0\\
                    Jilin     & 6 months  & 20km$\times$20km & 18k  & 665    & 261k   & 399k & 123 & 13.8\\
    			\bottomrule
    		\end{tabular}
     		\label{tab:pickup_data_statistics}
    	}
	}
	\label{datasets} 
\end{table}

\subsection{Dataset Properties \& Challenges} \label{sec:data_properties}
\par In this subsection, we present our primary data analysis to highlight its properties and the challenges they entail. 

\par \textbf{Large scale}. \texttt{LaDe} contains in total {10,667k} packages and {619k} trajectories that consist of {16,755k} GPS pings generated by {21k} couriers, covering 5 cities over a total span of 6 months. The maximal package number of a courier one trip in the pick-up scenario and delivery scenario reaches {95} and {121}, respectively. \textit{Such large scale brings a significant challenge to algorithms in last-mile delivery.} To the best of our knowledge, this is the largest clean delivery dataset available to the research community, in terms of spatio-temporal coverage, the total number of packages, and the number of couriers' trajectories. 

\par \textbf{Comprehensity}. \texttt{LaDe} aims to offer a wealth of information pertaining to last-mile delivery, encompassing various types of data such as detailed package information, task-event logs, courier trajectory details, and contextual features. The objective is to facilitate a broader range of research endeavors. \textit{How to effectively leverage these comprehensive features to improve existing or inspire new tasks remains an open problem for researchers from different communities.}

\par
\begin{figure}[htbp]
    \vspace{-0.5em}
    \centering
    \includegraphics[width=1 \linewidth]{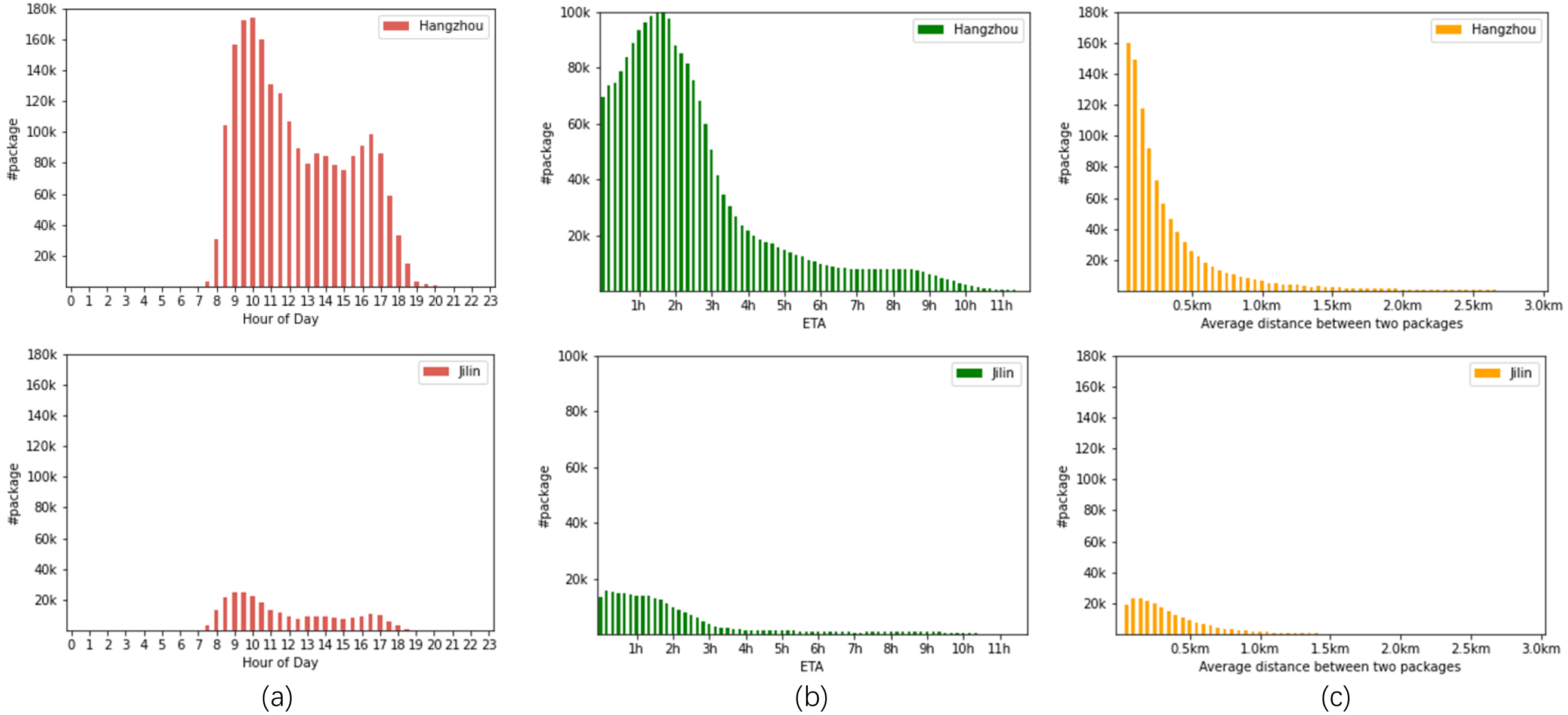}
    \caption{Diversity of cities. We select two cities, Hangzhou and Jilin, as an example to reveal their different spatio-temporal distributions. (a) The time distribution of packages in a day; (b) The ETA distribution of packages; (c) The distribution of the average distance between two consecutive packages in a courier's route. A significant difference is observed in the above illustration.}
\label{fig:different_cite_patterns}
\vspace{-0.5em}
\end{figure}

 \textbf{Diversity}. We increase the data's diversity from two perspectives: (1) scenario diversity -- we introduce scenario diversity by collecting two sub-datasets representing both pick-up and delivery scenarios; (2) city diversity -- we collect data from different cities to increase the diversity of the dataset. The cities in the dataset have different characteristics, leading to various spatio-temporal patterns in the dataset, where we give an illustration in Figure~\ref{fig:different_cite_patterns}. For more information about the selected cities, please refer to Table~\ref{tab:city_information} in Appendix~\ref{appendix:data_detail}. \textit{Such diversity brings the challenge of designing advanced models that can generalize well under cities with different characteristics.}


\par \textbf{Dynamism} (only for \texttt{LaDe-P}). Compared to \texttt{LaDe-D}, the tasks of a courier in \texttt{LaDe-P} are not settled at the beginning of the day. Rather, they are revealed along with the pick-up process as customers can place an order at any time. \textit{Such dynamism in courier tasks poses significant technical challenges in various research areas}, with one notable example being dynamic route optimization \cite{yao2019robust, li2021learning}.

\par Eqquiped with the above unique properties, \texttt{LaDe} offers the most extensive compilation of data for various research purposes background by last-mile delivery. It encompasses a variety of information across multiple domains, such as package details, event-based information, and courier information. Our aspiration is to make this abundant resource accessible to a broad spectrum of researchers, enabling them to undertake diverse and innovative studies.

\vspace{-0.5em}
\section{Applications} \label{sec:application}
\vspace{-0.5em}
\par
\begin{figure}[htbp]
	\centering
        \includegraphics[width=1\linewidth ]{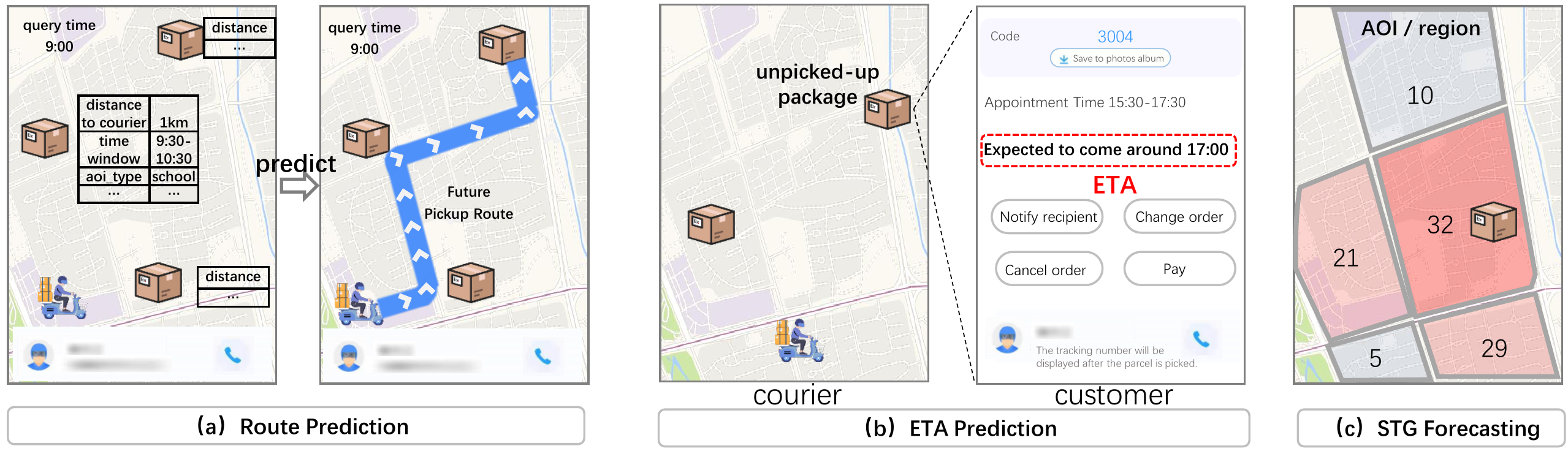}
	\caption{Illustration of three real-world applications. (a): Route prediction predicts the future pick-up route of a courier. (b): ETA prediction estimates the courier's arrival time for picking up or delivering packages. (c): STG forecasting predicts the future package number in given regions/AOIs.}
	\label{fig:application}
        \vspace{-1 em}
\end{figure}

 To prove \texttt{LaDe}'s ability to support multiple tasks, we benchmark the dataset in three learning-based tasks, including route prediction, estimated time of arrival prediction, and spatio-temporal graph forecasting. Those tasks all come from the real-world application and we illustrate them in Figure~\ref{fig:application}. The code of relevant baselines in each task is released at {https://huggingface.co/datasets/Cainiao-AI/LaDe}.  Note that the dataset can support far more than the three tasks, which we envision more possible applications from different research fields at the end of the section.  

\subsection{Route Prediction}
\par  A crucial task in last-mile delivery services (such as  logistics and food delivery) is service route prediction  \cite{gao2021deep, wen2022graph2route}, which aims to estimate the future service route of a worker given his unfinished tasks at the request time. 

\par \textbf{Problem Definition.} Formally, at a certain time $t$, a worker (i.e., courier) $w$ can have $n$ unfinished tasks, denoted by $\mathbf{X}_t^{w}=\{\mathbf{x}_1, \mathbf{x}_2, \dots, \mathbf{x}_n\}$, where $\mathbf{x}_i$ corresponds to the feature vector of a task $i$. Given a worker $w$'s unfinished tasks at time $t$ and route constraints $\mathcal C$ (such as pick-up then delivery constraints), route prediction aims to learn a mapping function ${\mathcal F}_{{\mathcal C}}$ to predict the worker's future service route $\hat {\bm \pi}$ which can satisfy the given route constraints $\mathcal C$, formulated as: 
$ {{\mathcal F}_{\mathcal C}}(\mathbf{X}_t^{w}) = [{\pi _1},{\pi _2} \cdots {\pi _{n}}]$, where $\pi_i$ means that the $i$-th node in the route is task ${\pi_i}$. And ${\pi _i} \in \{ 1, \cdots n\} \; {\rm and}\;{\pi _i} \ne {\pi _j}\;{\rm if}\;i \ne j$. 

\par \textbf{Dataset.} We choose \texttt{LaDe-P} as the dataset to conduct the experiment. The training, validation, and test set is split chronologically using a ratio of 6:2:2. Due to the space limit, we select three out of the five cities for conducting experiments, including Shanghai, Chongqing, and Yantai.

\par \textbf{Baselines \& Hyperparameters.} We run six baselines on \texttt{LaDe}. 1) Basic methods: TimeGreedy \cite{e_le_me} and DistanceGreedy \cite{e_le_me}. 2) Machine learning method: Osqure \cite{e_le_me}. 3) Deep learning models: DeepRoute \cite{wen2021package}, FDNET \cite{gao2021deep}, and Graph2Route \cite{wen2022graph2route}. Hyperparameters search is performed on the validation set by evaluating hidden size in \{16, 32, 64, 128\}. We set the learning rate to 0.0001 and batch size to 64 for all deep-learning models.  More details about the baselines and metrics can be found in Appendix~\ref{appendix:route_prediction_experiment}.

\par \textbf{Results}. Following \cite{wen2022graph2route}, we adopt HR@$k$, KRC, LMD, and ED to evaluate model performance. Higher KRC, HR@$k$, and lower LSD and ED mean better performance. The number of packages in each sample is in $(0, 25]$. Table \ref{tab:route_prediction_result} shows the results of different methods on \texttt{LaDe}. It can be observed that basic models perform poorly since they can only make use of distance or time information. Deep models generally achieve better performance than shallow models, because of their ability to model abundant spatial and temporal features. This further proves the importance of the comprehensive information provided by \texttt{LaDe} for building more powerful models. Among deep models, Graph2Route performs the best due to its ability to model the underlying graph correlation of different packages. 

\begin{table*}[htbp]
\centering
\small
\caption{\centering {Experiment Results of Route Prediction. }}
\renewcommand\arraystretch{1.1}
\setlength\tabcolsep{2 pt}
\resizebox{1.0 \textwidth}{!}{
\begin{tabular}{c|cccc|cccc|cccc}
\hline 
\multirow{2}{*}{Method}
& \multicolumn{4}{|c|}{Chongqing}& \multicolumn{4}{|c|}{Shanghai}& \multicolumn{4}{|c}{Yantai}\\
\cline{2-13}
 & HR@3 & KRC & LSD & ED & HR@3 & KRC & LSD & ED & HR@3 & KRC & LSD & ED\\
\hline \hline
Time-Greedy & {63.86~\tiny{$\pm$0.00}} & {44.16~\tiny{$\pm$0.00}} & {3.91~\tiny{$\pm${0.00}}} & {1.74~\tiny{$\pm${0.00}}} & {59.81~\tiny{$\pm$0.00}} & {39.93~\tiny{$\pm$0.00}} & {5.20~\tiny{$\pm${0.00}}} & {2.24~\tiny{$\pm${0.00}}} & {61.23~\tiny{$\pm$0.00}} & {39.64~\tiny{$\pm$0.00}} & {4.62~\tiny{$\pm${0.00}}} & {1.85~\tiny{$\pm${0.00}}} \\ \hline
Distance-Greedy & {62.99~\tiny{$\pm$0.00}} & {41.48~\tiny{$\pm$0.00}} & {4.22~\tiny{$\pm${0.00}}} & {1.60~\tiny{$\pm${0.00}}} & {61.07~\tiny{$\pm$0.00}} & {42.84~\tiny{$\pm$0.00}} & {5.35~\tiny{$\pm$\underline{0.00}}} & {1.94~\tiny{$\pm${0.00}}} & {62.34~\tiny{$\pm$0.00}} & {40.82~\tiny{$\pm$0.00}} & {4.49~\tiny{$\pm${0.00}}} & {1.64~\tiny{$\pm${0.00}}} \\ \hline
Or-Tools & {64.19~\tiny{$\pm$0.00}} & {43.09~\tiny{$\pm$0.00}} & {3.67~\tiny{$\pm$0.00}} & {1.55~\tiny{$\pm$0.00}} & {62.50~\tiny{$\pm$0.00}} & {44.81~\tiny{$\pm$0.00}} & {4.69~\tiny{$\pm$0.00}} & {1.88~\tiny{$\pm$0.00}} & {63.27~\tiny{$\pm$0.00}} & {42.31~\tiny{$\pm$0.00}} & {3.94~\tiny{$\pm$0.00}} & {1.59~\tiny{$\pm$0.00}} \\ \hline
LightGBM & {71.55~\tiny{$\pm$0.00}} & {54.53~\tiny{$\pm$0.00}} & {2.63~\tiny{$\pm$0.00}} & {1.54~\tiny{$\pm$0.00}} & {70.63~\tiny{$\pm$0.00}} & {54.48~\tiny{$\pm$0.00}} & {3.27~\tiny{$\pm$0.00}} & {1.92~\tiny{$\pm$0.00}} & {70.41~\tiny{$\pm$0.00}} & {52.90~\tiny{$\pm$0.00}} & {2.87~\tiny{$\pm$0.00}} & {1.59~\tiny{$\pm$0.00}} \\ \hline
FDNET & {69.98~\tiny{$\pm${0.32}}} & {52.07~\tiny{$\pm${0.38}}} & {3.36~\tiny{$\pm$0.04}} & {1.51~\tiny{$\pm$0.01}} & {69.05~\tiny{$\pm${1.23}}} & {52.72~\tiny{$\pm${1.72}}} & {4.08~\tiny{$\pm$0.25}} & {1.86~\tiny{$\pm$0.03}} & {69.08~\tiny{$\pm${0.61}}} & {50.62~\tiny{$\pm${1.20}}} & {3.60~\tiny{$\pm$0.15}} & {1.57~\tiny{$\pm$0.02}} \\ \hline
DeepRoute & {72.09~\tiny{$\pm${0.39}}} & {55.72~\tiny{$\pm${0.40}}} & {2.66~\tiny{$\pm$0.06}} & {1.51~\tiny{$\pm$0.01}} & {71.66~\tiny{$\pm$0.10}} & {56.20~\tiny{$\pm${0.23}}} & {3.26~\tiny{$\pm$0.07}} & {1.86~\tiny{$\pm$0.01}} & {71.44~\tiny{$\pm${0.28}}} & {54.74~\tiny{$\pm$\underline{0.49}}} & {2.80~\tiny{$\pm$0.02}} & {\underline{1.53}~\tiny{$\pm$0.02}} \\ \hline
CPRoute & {\underline{72.55}~\tiny{$\pm$0.23}} & {55.76~\tiny{$\pm$0.32}} & {2.70~\tiny{$\pm$0.01}} & {\underline{1.49}~\tiny{$\pm$0.02}} & {\underline{71.73}~\tiny{$\pm$0.08}} & {56.17~\tiny{$\pm$0.04}} & {3.39~\tiny{$\pm$0.02}} & {\underline{1.84}~\tiny{$\pm$0.00}} & {\underline{71.76}~\tiny{$\pm$0.04}} & {54.84~\tiny{$\pm$0.03}} & {2.99~\tiny{$\pm$0.01}} & {1.53~\tiny{$\pm$0.00}} \\ \hline
Graph2Route & {72.31~\tiny{$\pm$0.20}} & {56.08~\tiny{$\pm$0.14}} & {\underline{2.53}~\tiny{$\pm$0.01}} & {1.50~\tiny{$\pm$0.01}} & {71.69~\tiny{$\pm$0.10}} & {\underline{56.53}~\tiny{$\pm$0.10}} & {\underline{3.12}~\tiny{$\pm$0.01}} & {1.86~\tiny{$\pm$0.00}} & {71.52~\tiny{$\pm$0.14}} & {\underline{55.02}~\tiny{$\pm$0.10}} & {\underline{2.71}~\tiny{$\pm$0.01}} & {1.54~\tiny{$\pm$0.00}} \\ \hline
M2G4RTP & {72.44~\tiny{$\pm$0.11}} & {\underline{56.15}~\tiny{$\pm$0.04}} & {2.55~\tiny{$\pm$0.03}} & {1.50~\tiny{$\pm$0.00}} & {71.73~\tiny{$\pm$0.06}} & {56.34~\tiny{$\pm$0.08}} & {3.16~\tiny{$\pm$0.02}} & {1.86~\tiny{$\pm$0.01}} & {71.73~\tiny{$\pm$0.01}} & {55.02~\tiny{$\pm$0.17}} & {2.78~\tiny{$\pm$0.00}} & {1.53~\tiny{$\pm$0.01}} \\ \hline
DRL4Route & {\textbf{73.12}~\tiny{$\pm$0.06}} & {\textbf{57.23}~\tiny{$\pm$0.12}} & {\textbf{2.43}~\tiny{$\pm$0.01}} & {\textbf{1.48}~\tiny{$\pm$0.01}} & {\textbf{72.18}~\tiny{$\pm${0.15}}} & {\textbf{57.20}~\tiny{$\pm$0.18}} & {\textbf{3.06}~\tiny{$\pm$0.02}} & {\textbf{1.84}~\tiny{$\pm$0.01}} & {\textbf{72.07}~\tiny{$\pm$0.06}} & {\textbf{55.94}~\tiny{$\pm$0.10}} & {\textbf{2.62}~\tiny{$\pm$0.00}} & {\textbf{1.51}~\tiny{$\pm$0.00}} \\ \hline

\end{tabular}
}
\vspace{-10pt}
\label{tab:route_prediction_result}
\end{table*}

\subsection{Estimated Time of Arrival Prediction}
\par Estimated Time of Arrival (ETA) prediction aims to forecast when the task is going to be finished, e.g., the delivery time of a package. It is one of the most important tasks in many delivery platforms since it directly influences customers' experience \cite{wu2019deepeta}.

\par \textbf{Problem Definition.} Given an ETA query of worker $w$ at time $t$, i.e., $q = \{t, \mathbf{X}_t^{w}\}$, where $\mathbf{X}_t^{w}=\{\mathbf{x}_1, \mathbf{x}_2, \dots, \mathbf{x}_n\}$ is the courier's unfinished packages,  ETA prediction aims to build a model $ \mathcal{F}$ that can map the input query to the  arrival time (i.e., pick-up/delivery time) $\bm Y$ for the unfinished package set: $\mathcal{F}(q) \mapsto \bm Y = \{y_1, \dots, y_n\}$, where  ${y}_{i}= t^{\rm{actual}}_i - t$ and $t^{\rm{actual}}$ is the actual arrival time of task $i$.

\par \textbf{Dataset.} \texttt{LaDe-D} is utilized for this experiment (note that \texttt{LaDe-P} can also be used for this task). We split the data into  training, validation, and test sets chronologically in a ratio of 6:2:2. 

\par \textbf{Baselines \& Hyperparameters.}  Five baselines are evaluated for the task, including a simple speed-based method SPEED, machine learning methods LightGBM \cite{ke2017lightgbm} and KNN \cite{song2019service}, and deep models Multi-Layer Perceptron (MLP) and FDNET \cite{gao2021deep}. We also perform hyperparameters search on the validation set by hidden size in \{16, 32, 64, 128\} for all deep models. The learning rate and batch size are set to 0.00005 and 32 for all models. See more details in Appendix~\ref{appendix:time_prediction_experiment}.


\par \textbf{Results.} MAE, RMSE, and ACC@20($\%$) are used to evaluate the performance of time prediction models. Higher ACC@20 and lower MAE and RMSE indicate better performance. From the results shown in Table \ref{tab:time_prediction_result}, we can see that learning-based models outperform SPEED by a large margin because of their ability to model multiple spatio-temporal factors. We also observe a huge performance gap of the same method in different cities. For example, the best model, RANKETPA, achieves 73.67\% in terms of ACC@20 in Shanghai, while it gets a much lower accuracy of 57.83\% and 54.45\% in the other two datasets. It deserves further study to build a more powerful model that can generalize well in cities with different properties.

\begin{table*}[htbp]
\centering
\small
\caption{\centering {Experiment results of ETA prediction. }}
\renewcommand\arraystretch{1.1}
\setlength\tabcolsep{2 pt}
\resizebox{1.0 \textwidth}{!}{
\begin{tabular}{c|ccc|ccc|ccc}
\hline 
{\multirow{2}{*}{Method}}& \multicolumn{3}{|c|}{Shanghai}& \multicolumn{3}{|c|}{Yantai}& \multicolumn{3}{|c}{Chongqing}\\
\cline{2-10}
 & MAE $\downarrow$ & RMSE $\downarrow$ & ACC@20 $\uparrow$ & MAE $\downarrow$ & RMSE $\downarrow$ & ACC@20 $\uparrow$ & MAE $\downarrow$ & RMSE $\downarrow$ & ACC@20 $\uparrow$\\
\hline \hline
SPEED & {26.68~\tiny{$\pm${0.00}}} & {31.31~\tiny{$\pm${0.00}}} & {52.57~\tiny{$\pm$0.00}} & {33.97~\tiny{$\pm${0.00}}} & {40.27~\tiny{$\pm${0.00}}} & {42.03~\tiny{$\pm$0.00}} & {35.55~\tiny{$\pm${0.00}}} & {42.06~\tiny{$\pm${0.00}}} & {41.10~\tiny{$\pm$0.00}} \\ \hline
KNN & {25.22~\tiny{$\pm${0.00}}} & {29.57~\tiny{$\pm${0.00}}} & {65.71~\tiny{$\pm$0.00}} & {28.10~\tiny{$\pm${0.00}}} & {33.80~\tiny{$\pm${0.00}}} & {45.33~\tiny{$\pm$0.00}} & {29.45~\tiny{$\pm${0.00}}} & {35.19~\tiny{$\pm${0.00}}} & {43.68~\tiny{$\pm$0.00}} \\ \hline
LightGBM & {17.24~\tiny{$\pm$0.00}} & {20.40~\tiny{$\pm$0.00}} & {67.44~\tiny{$\pm$0.00}} & {23.32~\tiny{$\pm$0.00}} & {27.82~\tiny{$\pm$0.00}} & {51.22~\tiny{$\pm$0.00}} & {24.22~\tiny{$\pm$0.00}} & {27.99~\tiny{$\pm$0.00}} & {48.80~\tiny{$\pm$0.00}} \\ \hline
MLP & {{16.16}~\tiny{$\pm$0.02}} & {{19.31}~\tiny{$\pm$0.01}} & {{72.17}~\tiny{$\pm$0.10}} & {{22.18}~\tiny{$\pm$0.06}} & {26.38~\tiny{$\pm$0.06}} & {{54.98}~\tiny{$\pm${0.17}}} & {23.82~\tiny{$\pm$0.02}} & {28.24~\tiny{$\pm$0.02}} & {{52.74}~\tiny{$\pm${0.16}}} \\ \hline
FDNET & {18.81~\tiny{$\pm$1.23}} & {21.15~\tiny{$\pm$2.47}} & {64.30~\tiny{$\pm${1.43}}} & {22.41~\tiny{$\pm$0.50}} & {{26.00}~\tiny{$\pm$0.63}} & {54.67~\tiny{$\pm${1.55}}} & {\underline{22.54}~\tiny{$\pm$0.91}} & {\underline{24.53}~\tiny{$\pm$0.92}} & {46.65~\tiny{$\pm${4.82}}} \\ \hline

RankETPA & {\underline{15.76}~\tiny{$\pm$0.05}} & {\underline{19.13}~\tiny{$\pm$0.08}} & {\underline{73.67}}~\tiny{$\pm${0.15}} & {\underline{21.36}~\tiny{$\pm$0.07}} & {\underline{25.52}~\tiny{$\pm$0.08}} & {\underline{57.83}~\tiny{$\pm$0.07}} & {{23.37}~\tiny{$\pm$0.01}} & {{27.93}~\tiny{$\pm$0.04}} & {\underline{54.45}~\tiny{$\pm$0.10}} \\ \hline

M2G4RTP & {\textbf{8.23}~\tiny{$\pm$0.02}} & {\textbf{9.59}~\tiny{$\pm$0.03}} & {\textbf{91.01}}~\tiny{$\pm${0.51}} & {\textbf{17.21}~\tiny{$\pm$1.57}} & {\textbf{19.71}~\tiny{$\pm$1.88}} & {\textbf{69.20}~\tiny{$\pm$0.39}} & {\textbf{17.96}~\tiny{$\pm$0.25}} & {\textbf{19.74}~\tiny{$\pm$0.31}} & {\textbf{65.97}~\tiny{$\pm$0.32}} \\ \hline
\end{tabular}
}
\vspace{-10pt}
\label{tab:time_prediction_result}
\end{table*}

\subsection{Spatio-Temporal Graph (STG) Forecasting} 

\par \texttt{LaDe} contains the package data with information that records when and where the package order is placed. Based on this, the package number of a region within a certain period can be calculated. In this way, \texttt{LaDe} also contributes as a new dataset to another well-known task -- \emph{spatio-temporal graph forecasting} \cite{li2018diffusion, yao2018deep, simeunovic2021spatio}, which aims to predict future graph signals given its historical observations.

\par \textbf{Problem Definition.} Let $\mathcal{G}=\{{\mathcal V}, {\mathcal E}, {\mathbf A} \}$ represent a graph with $V$ nodes, where $\mathcal V$,  $\mathcal E$ are the node set and edge set, respectively. ${\mathbf A} \in {\mathbb R}^{V \times V}$ is a weighted adjacency matrix to describe the graph topology. 
For ${\mathcal V}=\{ v_1, \dots, v_{V} \}$, let ${\mathbf x}_t \in {\mathbb R}^{ F \times V}$ denote $F$-dimentional signals generated by the $V$ nodes at time $t$. Given historical graph signals $\xh = [{\mathbf x_1}, \cdots, {\mathbf x}_{T_h}]$ of $T_h$  time steps and the graph $\G$ as inputs, STG forcasting aims at learning a function $\mathcal F$ to predict future graph signals $\xp{}$, formulated as: $ {\mathcal F}:(\xh; \G) \rightarrow [{\mathbf x_{T_h + 1}}, \cdots, {\mathbf x_{T_h + T_p}}]:=\xp{}$, where $T_p$ is the forecasting horizon.

\par \textbf{Dataset.} \texttt{LaDe-P} is used to conduct this experiment. More experiment details can be found in Appendix~\ref{appendix:stg_prediction_experiment}. Each node corresponds to a region within the city. The signal of each node represents the number of packages picked up during a particular time stamp. We set the time interval to be 1 hour. Our objective is to leverage the data from the previous 24 hours to predict the package volume for the subsequent 24 hours. We use the ratio of 6:2:2 for training, evaluation, and testing sets  based on the chronological order of the timestamps.

\par \textbf{Baselines  \& Hyperparameters.} We evaluate eight baselines, including a traditional method (i.e., HA \cite{ha}), and recent deep learning models, including DCRNN \cite{li2018diffusion}, STGCN \cite{stgcn}, GWNET \cite{gwnet}, ASTGCN \cite{guo2019attention}, MTGNN \cite{mtgnn}, AGCRN \cite{bai2020adaptive} and STGNCDE \cite{stgncde}. We set the hidden size, learning rate, and batch size to 32, 0.001, and 32 for all models.
 

\par \textbf{Results.} MAE and RMSE are used as the metrics, and results are shown in Table~\ref{tab:st_results}. Comparing the methods, we can observe that the baseline method, HA (Historical Average), achieves relatively higher MAE and RMSE values across all three cities. This indicates that simply using historical averages to predict future spatio-temporal graph data is not as effective as the other methods. The results of the different methods may vary slightly depending on the city. For instance, in Shanghai, GWNET, ASTGCN, and MTGNN exhibit similar performance, while in Hangzhou and Chongqing, MTGNN and ASTGCN achieve the lowest errors, respectively.


\begin{table*}[htbp]
\centering
\footnotesize
\caption{\centering {Experimental results of spatio-temporal graph prediction.}}
\setlength\tabcolsep{8 pt}
\resizebox{1.0 \textwidth}{!}{
\begin{tabular}{l|cc|cc|cc}
\hline
\multicolumn{1}{c|}{\multirow{2}{*}{Method}} & \multicolumn{2}{c|}{Shanghai} & \multicolumn{2}{c|}{Hangzhou} & \multicolumn{2}{c}{Chongqing}   \\
\cline{2-7}
\multicolumn{1}{c|}{}                        & MAE $\downarrow$          & RMSE $\downarrow$        & MAE  $\downarrow$         & RMSE $\downarrow$         & MAE $\downarrow$         & RMSE  $\downarrow$           \\\hline \hline
 HA \cite{ha}                   & $4.63$          & $9.91$         & $4.78$          & $10.53$         & $2.44$          & $5.30$          \\
 DCRNN \cite{li2018diffusion}   & 3.69 \tiny{$\pm$ 0.09} & 7.08 \tiny{$\pm$ 0.12} & 4.14 \tiny{$\pm$ 0.02} & 7.35 \tiny{$\pm$ 0.07} & 2.75 \tiny{$\pm$ 0.07} & 5.11 \tiny{$\pm$ 0.12} \\
 STGCN \cite{stgcn}             & \textbf{3.04 \tiny{$\pm$ 0.02}} & \textbf{6.42 \tiny{$\pm$ 0.05}} & \underline{3.01 \tiny{$\pm$ 0.04}} & \underline{5.98 \tiny{$\pm$ 0.10}} & 2.16 \tiny{$\pm$ 0.01} & 4.38 \tiny{$\pm$ 0.03} \\
 GWNET \cite{gwnet}             & 3.16 \tiny{$\pm$ 0.06} & 6.56 \tiny{$\pm$ 0.11} & 3.22 \tiny{$\pm$ 0.03} & 6.32 \tiny{$\pm$ 0.04} & 2.22 \tiny{$\pm$ 0.03} & 4.45 \tiny{$\pm$ 0.05} \\
 ASTGCN \cite{guo2019attention} & \underline{3.12 \tiny{$\pm$ 0.06}} & \underline{6.48 \tiny{$\pm$ 0.14}} & 3.09 \tiny{$\pm$ 0.04} & 6.06 \tiny{$\pm$ 0.10} & \textbf{2.11 \tiny{$\pm$ 0.02}} & \textbf{4.24 \tiny{$\pm$ 0.03}} \\
 MTGNN \cite{mtgnn}             & 3.13 \tiny{$\pm$ 0.04} & 6.51 \tiny{$\pm$ 0.13} & \textbf{3.01 \tiny{$\pm$ 0.01}} & \textbf{5.83 \tiny{$\pm$ 0.03}} & \underline{2.15 \tiny{$\pm$ 0.01}} & \underline{4.28 \tiny{$\pm$ 0.05}} \\
 AGCRN \cite{bai2020adaptive}   & 3.93 \tiny{$\pm$ 0.03} & 7.99 \tiny{$\pm$ 0.08} & 4.00 \tiny{$\pm$ 0.03} & 7.88 \tiny{$\pm$ 0.06} & 2.46 \tiny{$\pm$ 0.00} & 4.87 \tiny{$\pm$ 0.01} \\
 STGNCDE \cite{stgncde}         & 3.74 \tiny{$\pm$ 0.15} & 7.27 \tiny{$\pm$ 0.16} & 3.55 \tiny{$\pm$ 0.04} & 6.88 \tiny{$\pm$ 0.10} & 2.32 \tiny{$\pm$ 0.07} & 4.52 \tiny{$\pm$ 0.07} \\
\hline
\end{tabular}
}
\label{tab:st_results}
\vspace{-0.5em}
\end{table*}

\subsection{Disscussion of Other Potential Tasks} \label{sec:disscuss_of_other_tasks}
\par In addition to primary tasks, the dataset can provide substantial support for a wide range of other tasks within the context of last-mile delivery. In the future, we plan to explore a wider range of applications on \texttt{LaDe}. 
Here we present a list of tasks supported by \texttt{LaDe} in Table~\ref{tab:task_information}, highlighting the minimal required information necessary for performing each task using \texttt{LaDe}. This effectively showcases \texttt{LaDe}'s remarkable multi-task support capability.


\begin{table}[htbp]
	\centering
	\caption{Supported tasks with the minimal required information.}
	\setlength\tabcolsep{6 pt}
	\resizebox{1 \linewidth}{!}
	{
		\begin{tabular}{l|cccccc}
			\hline
		      Task & Package Info & Stop Info & Courier Info & Task-event Info & Context  \\
                \hline
                
                Vehicle Routing \cite{zeng2019last} & $\checkmark$ & $\checkmark$ &   & $\checkmark$ &   \\
                 Delivery Scheduling \cite{han2017appointment} & $\checkmark$ & $\checkmark$ &   & $\checkmark$ &   \\
                Last-Mile Data Mining \cite{ji2019alleviating, ruan2022discovering} & $\checkmark$ & $\checkmark$ & $\checkmark$ & $\checkmark$ & $\checkmark$ \\ 
                 Spatial Crowdsourcing  \cite{chen2016crowddeliver, han2017appointment, chen2020crowdexpress} & $\checkmark$ & $\checkmark$ &  $\checkmark$  & $\checkmark$ &   \\
                Time Prediction \cite{ruan2020doing,ruan2022service} & $\checkmark$ & $\checkmark$ & $\checkmark$ & $\checkmark$ &  \\ 
                Route Prediction \cite{gao2021deep,wen2022graph2route} & $\checkmark$ & $\checkmark$ & $\checkmark$ & $\checkmark$ &   \\ 
                STG Forecasting \cite{yao2018deep, simeunovic2021spatio} & $\checkmark$ & $\checkmark$ &   &   &  $\checkmark$ \\ 
			\hline
		\end{tabular}
 		\label{tab:task_information}
	}
	\label{datasets}
 \vspace{-0.5em}
\end{table}



\subsection{Data Limitations} \label{sec:limitations}
\par In this subsection, we introduce two potential limitations associated with the utilization of \texttt{LaDe}. 
The first one is the limited country coverage. \texttt{LaDe} is collected from the last-mile delivery data in the Cainiao logistic platform, which majorly targets the cities in China. The second limitation arises from the absence of parts of the courier's trajectory point. In the actual operation process, the platform cannot locate the location of couriers due to the problems of PDA (personal digital device) or GPS locating problems. As a result, there are some missing values in the courier's location at the accept time and pick-up/delivery time. The missing location rate when accepting orders is about {40\%}, and the  missing rate of GPS  when  picking up is {29\%} in \texttt{LaDe-P}.


\section{Conclusion}
\par In this paper, we introduced \texttt{LaDe}, the first comprehensive industry-scale last-mile delivery dataset, addressing the lack of a widely accepted, publicly available dataset for last-mile delivery research. \texttt{LaDe} provides a critical resource for researchers and practitioners to develop advanced algorithms in the context of last-mile delivery, with its large-scale, comprehensive, diverse, and dynamic characteristics enabling it to serve as a new and challenging benchmark dataset. We have also demonstrated the versatility of \texttt{LaDe} by benchmarking it on three real-world tasks, showcasing its potential applications in various research fields. The source code is released along with the dataset to drive the development of this area. By releasing \texttt{LaDe}, we aim to promote further research and collaboration among researchers from different fields, encouraging them to utilize it for developing novel algorithms and models, as well as comparing and validating their methods against state-of-the-art approaches. We believe that \texttt{LaDe} will significantly contribute to ongoing efforts to improve efficiency, cost-effectiveness, and customer satisfaction in last-mile delivery, ultimately benefiting the research community and logistics industry.

\bibliographystyle{IEEEtran}
\bibliography{reference.bib}









\section*{Checklist}

The checklist follows the references.  Please
read the checklist guidelines carefully for information on how to answer these
questions.  For each question, change the default \answerTODO{} to \answerYes{},
\answerNo{}, or \answerNA{}.  You are strongly encouraged to include a {\bf
justification to your answer}, either by referencing the appropriate section of
your paper or providing a brief inline description.  For example:
\begin{itemize}
  \item Did you include the license to the code and datasets? \answerYes{See Section~\ref{gen_inst}.}
  \item Did you include the license to the code and datasets? \answerNo{The code and the data are proprietary.}
  \item Did you include the license to the code and datasets? \answerNA{}
\end{itemize}
Please do not modify the questions and only use the provided macros for your
answers.  Note that the Checklist section does not count towards the page
limit.  In your paper, please delete this instructions block and only keep the
Checklist section heading above along with the questions/answers below.

\begin{enumerate}

\item For all authors...
\begin{enumerate}
  \item Do the main claims made in the abstract and introduction accurately reflect the paper's contributions and scope?
    \answerYes{}
  \item Did you describe the limitations of your work?
    \answerYes{See Section~\ref{sec:limitations}.}
  \item Did you discuss any potential negative societal impacts of your work?
    \answerYes{See Section\ref{sec:data_collection}}
  \item Have you read the ethics review guidelines and ensured that your paper conforms to them?
    \answerYes{}
\end{enumerate}

\item If you are including theoretical results...
\begin{enumerate}
  \item Did you state the full set of assumptions of all theoretical results?
    \answerNA{}
	\item Did you include complete proofs of all theoretical results?
    \answerNA{}
\end{enumerate}

\item If you ran experiments (e.g. for benchmarks)...
\begin{enumerate}
  \item Did you include the code, data, and instructions needed to reproduce the main experimental results (either in the supplemental material or as a URL)?
    \answerYes{See Introduction.}
  \item Did you specify all the training details (e.g., data splits, hyperparameters, how they were chosen)?
    \answerYes{See Section~\ref{sec:application}.}
	\item Did you report error bars (e.g., with respect to the random seed after running experiments multiple times)?
    \answerYes{}
	\item Did you include the total amount of compute and the type of resources used (e.g., type of GPUs, internal cluster, or cloud provider)?
    \answerNA{}
\end{enumerate}

\item If you are using existing assets (e.g., code, data, models) or curating/releasing new assets...
\begin{enumerate}
  \item If your work uses existing assets, did you cite the creators?
    \answerYes{}
  \item Did you mention the license of the assets?
    \answerYes{See Section~\ref{sec:introduction}.}
  \item Did you include any new assets either in the supplemental material or as a URL?
    \answerYes{See Section~\ref{sec:introduction}}
  \item Did you discuss whether and how consent was obtained from people whose data you're using/curating?
    \answerYes{See Section~\ref{sec:data_collection}.}
  \item Did you discuss whether the data you are using/curating contains personally identifiable information or offensive content?
    \answerYes{See Section~\ref{sec:data_collection}.}
\end{enumerate}

\item If you used crowdsourcing or conducted research with human subjects...
\begin{enumerate}
  \item Did you include the full text of instructions given to participants and screenshots, if applicable?
    \answerNA{}
  \item Did you describe any potential participant risks, with links to Institutional Review Board (IRB) approvals, if applicable?
    \answerNA{}
  \item Did you include the estimated hourly wage paid to participants and the total amount spent on participant compensation?
    \answerNA{}
\end{enumerate}

\end{enumerate}


\newpage
\appendix

\par \textbf{Appendix}

\section{Detailed Dataset Description} 

\subsection{Data Field} \label{appendix:data_detail}

\begin{table}[htbp]
	\centering
	\caption{Description of data fields of \texttt{LaDe-P.}}
	\setlength\tabcolsep{3 pt}
	\resizebox{\linewidth}{!}
	{
		\begin{tabular}{ccc}
			\toprule
			Data field & Description & Unit/format \\
			\toprule
			\multicolumn{3}{c}{Package information} \\
			
			\midrule
			package\_id &  Unique identifier of each package  & Id \\
			
			time\_window\_start  & start of the required time window  & Time \\
			time\_window\_end  & end of the required time window  & Time \\
			
			\toprule
			\multicolumn{3}{c}{Stop information} \\
			\midrule
			lng/lat & Coordinates of each stop & Float\\
			city &  City   &   String \\
                region\_id &  Id of the Region & String\\
			aoi\_id & Id of the AOI (Area of Interest) & Id \\
			aoi\_type & Type of the AOI & Categorical \\

			\toprule
			\multicolumn{3}{c}{Courier Information} \\
			\midrule
			courier\_id &  Id of the courier  &   Id \\

			\toprule
			\multicolumn{3}{c}{Task-event Information} \\
			\midrule
			accept\_time & The time when the courier accepts the task & Time \\
			accept\_gps\_time & The time of the GPS point  whose time  is the closest to accept time  & Time \\
			
			accept\_gps\_lng/accept\_gps\_lat & Coordinates when the courier accept the task & Float\\ \\
			
			pickup\_time & The time when the courier picks up the task & Time \\
			
			pickup\_gps\_time & The time of the GPS point  whose time  is the closest to the pickup\_time  & Time\\
			pickup\_gps\_lng/got\_gps\_lat & Coordinates when the courier picks up the task & Float\\

			\toprule
			\multicolumn{3}{c}{Context information} \\
			\midrule
		    ds &  the date of the package pickup  &   Date \\
		
			\bottomrule
		\end{tabular}
 		\label{tab:pickup_data_field}
	}
	\label{datasets}
\end{table}

\begin{table}[htbp]
	\centering
	\caption{Description of data fields of \texttt{LaDe-D.}}
	\setlength\tabcolsep{3 pt}
	\resizebox{\linewidth}{!}
	{
		\begin{tabular}{ccc}
			\toprule
			Data field & Description & Unit/format \\
			\toprule
			\multicolumn{3}{c}{Package information} \\
			
			\midrule
			package\_id &  Unique identifier of each package  &   Id \\
			
			\toprule
			\multicolumn{3}{c}{Stop information} \\
			\midrule
			lng/lat & Coordinates of each stop  &  Float \\
			city &  City   &   String \\
                region\_id & Id of the region & Id \\
			aoi\_id & Id of the AOI & Id \\
			aoi\_type & Type of the AOI & Categorical \\

			\toprule
			\multicolumn{3}{c}{Courier Information} \\
			\midrule
			courier\_id &  Id of the courier  &   Id \\

			\toprule
			\multicolumn{3}{c}{Task-event Information} \\
			\midrule
			accept\_time & The time when the courier accept the task & Time \\
			accept\_gps\_time & The time of the GPS point  whose time  is the closest to accept time  & Time \\
			
			accept\_gps\_lng/accept\_gps\_lat & Coordinates when the courier accept the task & Float\\ \\
			
			delivery\_time & The time when courier finishes delivering the task & Time \\

			delivery\_gps\_time & The time of the GPS point  whose time  is the closest to the got time  & Time\\
			delivery\_gps\_lng/delivery\_gps\_lat & Coordinates when the courier finish the task & Float\\

			\toprule
			\multicolumn{3}{c}{Context information} \\
			\midrule
		    ds &  the date of the package delivery  &   Date \\
		
			\bottomrule
		\end{tabular}
 		\label{tab:delivery_data_field}
	}
	\label{datasets}
\end{table}

\begin{table}[htbp]
	\centering
	\caption{Description of data fields of \texttt{Detailed Trajectory.}}
	\setlength\tabcolsep{3 pt}
	\resizebox{0.7 \linewidth}{!}
	{
		\begin{tabular}{ccc}
			\toprule
			Data field & Description & Unit/format \\
			\midrule
			ds & The date of the trajectory  & Date \\
			courier\_id & Id of the courier  & Id \\
                gps\_time & The time when the trajectory point is recorded & Time \\
			lng/lat & Coordinates of the courier & Float\\ 

			\bottomrule
		\end{tabular}
 		\label{tab:trajectory_data}
	}
	\label{datasets}
\end{table}

\begin{table}[htbp]
	\centering
	\caption{Description of data fields of \texttt{Road Network.}}
	\setlength\tabcolsep{3 pt}
	\resizebox{\linewidth}{!}
	{
		\begin{tabular}{ccc}
			\toprule
			Data field & Description & Unit/format \\
			\midrule
			id & Unique identifier of the road  & Id \\
			road\_id & Id of a street or place name  & Id \\
                code & 4 digit code (between 1000 and 9999) defining the geographical features of a road & Id \\
	fclass & Class name of the road & String\\ 
   	ref & Reference number of the road & String\\ 
        oneway & Whether the road is a oneway road & String\\ 
        maxspeed & Max allowed speed in km/h & Int\\ 
        layer & Relative layering of roads & Int\\ 
        bridge & Whether the road is on a bridge & String\\ 
        tunnel & Whether the road is in a tunnel & String\\ 
        city & City of the road & String\\ 
        geometry & A geometry type composed of one or more line segments & String\\
                               
			\bottomrule
		\end{tabular}
 		\label{tab:trajectory_data}
	}
	\label{datasets}
\end{table}

\begin{table}[htbp]
	\centering
	\caption{Information of different selected cities.}
	\setlength\tabcolsep{2 pt}
	\resizebox{1 \linewidth}{!}
	{
		\begin{tabular}{c|cc}
			\toprule
		      City &  Description \\
                Shanghai &   One of the most prosperous cities in China, with a large number of orders per day. \\
                Hangzhou &   A big city with well-developed online e-commerce and a large number of orders per day.\\
                Chongqing &   A big city with complicated road conditions in China, with a large number of orders. \\
                Jilin &   A middle-size city in China, with a small number of orders each day.\\
                Yantai &   A small city in China, with a small number of orders every day. \\
			\bottomrule
		\end{tabular}
 		\label{tab:city_information}
	}
	\label{datasets}
\end{table}


\subsection{Data Statistics} \label{appendix:data_statistics}
\par Table~\ref{tab:delivery_data_statistics} shows the detailed statistics of \texttt{LaDe-D}.

\begin{table}[htbp]
	\centering
	\caption{Statistics of \texttt{LaDe-D}. AvgETA stands for the average arrival time per package. AvgPackage means the average package number of a courier per day. The unit of AvgETA is minute.}
	\setlength\tabcolsep{2 pt}
	\resizebox{\linewidth}{!}{
    	{
    		\begin{tabular}{cccccccccc}
    			\toprule
    			City & Time span & Spatial span & \#Trajectories & \#Couriers  & \#Packages &\#GPS points & AvgETA & AvgPackage\\
    			\midrule
    			Shanghai & 6 months & 20km$\times$20km & 70k & 1,733 & 1,483k & 2,967k & 102 & 21.1 & \\
                    Hangzhou & 6 months & 20km$\times$20km & 71k & 1,392 & 1,861k & 3,723k & 147 & 25.9 & \\
                    Chongqing & 6 months & 20km$\times$20km & 68k & 1,494 & 931k & 1,862k & 182 & 13.5 & \\
                    Yantai & 6 months & 20km$\times$20km & 17k & 205 & 206k & 410k & 244 & 11.5 & \\
                    Jilin & 6 months & 20km$\times$20km & 2k & 57 & 31k & 61k & 203 & 16.2 & \\
    			\bottomrule
    		\end{tabular}
     		\label{tab:delivery_data_statistics}
    	}
	}
	\label{datasets}
\end{table}

\begin{table}[htbp]
	\centering
	\caption{Statistics of \texttt{Detailed Trajectory}.}
	\setlength\tabcolsep{2 pt}
	\resizebox{0.7\linewidth}{!}{
    	{
    		\begin{tabular}{cccccccccc}
    			\toprule
    			City & Time span & Spatial span & \#GPS points & \#Couriers \\
    			\midrule
    			Shanghai & 1 month & 77km$\times$73km & 9727k & 245 & \\
                    Hangzhou & 1 months & 211km$\times$142km & 17900k & 443 & \\
                    Chongqing & 1 months & 403km$\times$267km & 13520k & 658 & \\
                    Yantai & 1 months & 155km$\times$114km & 6616k & 245  & \\
                    Jilin & 1 months & 105km$\times$165km & 2849k & 613 &  \\
    			\bottomrule
    		\end{tabular}
     		\label{tab:delivery_data_statistics}
    	}
	}
	\label{datasets}
\end{table}

\section{Experiments Details} \label{appendix:experiment_details}

\subsection{Experiment Details of Route Prediction} \label{appendix:route_prediction_experiment}
\textbf{Methods.} We adopt the following methods for experiments:
\begin{itemize}[leftmargin=*]
    \item TimeGreedy \cite{e_le_me}: A greedy algorithm, which ranks all the candidate tasks by sorting their remaining time.
    \item DistanceGreedy \cite{e_le_me}: A greedy algorithm, which chooses to take the nearest package at each step, regardless of time requirements and other factors.
    \item Osqure \cite{e_le_me}: A machine learning method, which predicts the next package at each time step through a machine learning algorithm,  by considering it as a multi-class classification problem.
    \item DeepRoute \cite{wen2021package}: A deep learning method, equipped with a Transformer encoder and Pointer Net decoder.
    \item FDNET \cite{gao2021deep}: A deep learning method, equipped with a Bi-LSTM encoder and Pointer Net decoder.
    \item Graph2Route \cite{wen2022graph2route}: A deep learning method, equipped with a dynamic graph encoder and personalized route decoder.
\end{itemize}

\textbf{Metrics.} Following the setting in \cite{wen2022graph2route}, the following metrics are utilized to evaluate the performance of route prediction methods:
\begin{itemize}[leftmargin=*]
\item  \textbf{KRC}: Kendall Rank Correlation \cite{kendall1938new} is a statistical metric to measure the ordinal association between two sequences.
Let $\hat{Y}$ and $Y$ be two sequences and $R_{\hat{Y}}(i) \in [1,|Y|]$ be the position of item $i$ in $Y$, a node pair $(i, j)$ is said to be concordant if and only if both $R_{\hat{Y}}(i) > R_{\hat{Y}}(j)$ and $R_{Y}(i) > R_{Y}(j)$, or both $R_{\hat{Y}}(i) < R_{\hat{Y}}(j)$ and $R_{Y}(i) < R_{Y}(j)$. Otherwise, it is said to be discordant.
To calculate this metric, nodes in the prediction are first divided into two sets: i) nodes in label ${{\mathcal V}_{in}} = \{ {\hat y}_i | {\hat y}_i \in {Y} \}$, and ii) nodes not in label ${{\mathcal V}_{not}} = \{ {\hat y}_i |  {\hat y}_i \not \in {Y} \}$. The order of items in $\mathcal{V}_{in}$ is available, while it is hard to tell the order of items in  ${\mathcal V}_{not}$. Still, we know that all items in $\mathcal{V}_{in}$ are ahead of that in ${\mathcal V}_{not}$. Therefore, we compare the nodes pairs $\{(i,j) | i,j \in {\mathcal V}_{in}~{\rm and}~{i \neq j}\} \cup \{(i,j) | i \in {\mathcal V}_{in} {~\rm and~~} j \in {\mathcal V}_{not} \}$. To this end, KRC is defined as:

\begin{equation}
	{\rm{KRC}} = \frac{N_c-N_d}{N_c+N_d},
	\label{eq:krc}
\end{equation}
where $N_c$ is the number of concordant pairs, and $N_d$ is the number of discordant pairs.
		
\item  \textbf{ED:} Edit Distance \cite{nerbonne1999edit} (ED) is an indicator to quantify how dissimilar two sequences $Y$ and $\hat{Y}$ are to one another, by counting the minimum number of required operations to transform one sequence into another. 

\item   \textbf{LSD}: Location Square Deviation (LSD) measures the degree that the prediction deviates from the label, formulated as:
\begin{equation}
    {\rm LMD} =\frac{1}{m}\sum_{i=1}^{m}|(R_{Y}(i)-R_{\hat{Y}}(i))|. 
    \label{eq:lsd_lmd}
\end{equation}

\item   \textbf{HR@$k$}: Hit-Rate@$k$  quantifies the similarity between the top-$k$ items of two sequences. It describes how many of the first $k$ predictions are in the label, which is formulated as follows:
\begin{equation}
    \textbf{\rm HR@}k=\frac{{\hat {Y}}_{[1:k]}\cap {Y}_{[1:k]}}{k}.
    \label{eq_hit_rate}
\end{equation}
\end{itemize}

\subsection{Experiment Details of Time Prediction} \label{appendix:time_prediction_experiment}
\textbf{Methods.} The following methods are chosen for experiments:
\begin{itemize}[leftmargin=*]
    \item SPEED, a simple speed-based method that utilizes distance/speed as the prediction value, where speed is calculated based on each worker's history trajectories. We set the speed for workers without previous trajectories as the average speed calculated by all workers.
    \item LightGBM \cite{ke2017lightgbm}, a popular machine-learning method for regression tasks.
    \item KNN \cite{song2019service}, a machine-learning method that trains a regressor based on K-Nearest Neighbors algorithm to predict the arrival time.
    \item MLP \cite{popescu2009multilayer}, a deep neural network model with 2 layers of MLPs.
    \item FDNet \cite{gao2021deep}, a deep model that predicts both route and time of unfinished tasks.
    
\end{itemize}

\textbf{Metrics.} MAE (Mean Absolute Error) and RMSE (Root Mean Squared Error), and ACC@30 are utilized as metrics. Note that delivery platforms usually provide an interval of arrival time for customer notification. Thus we compute the ratio of prediction where the time difference between predicted time and true time is less than $30$ minutes (ACC@30), formulated as $ \textbf{\rm ACC@30}= \frac{1}{N}\sum_{i=1}^N{\mathbb{I}}(|\hat{y_i}-y_i|<30).$

\subsection{Experiment Details of Spatio-temporal Graph Forecasting} \label{appendix:stg_prediction_experiment}

\textbf{Methods.} For our Spatio-temporal Graph Forecasting experimental setup, we have selected the following methods:

\begin{itemize}[leftmargin=*]

    \item \textbf{HA} \cite{ha}: HA predicts future values of a time series by calculating the mean of past observations that correspond to the same time periods.

    
\item \textbf{DCRNN} \cite{li2018diffusion}: DCRNN employs a neural network architecture that incorporates diffusion convolution and sequence-to-sequence mechanisms. This enables the model to effectively learn spatial dependencies and temporal relations within the data.

\item \textbf{STGCN} \cite{stgcn}: STGCN is a specialized spatio-temporal graph convolution network that synergistically merges spectral graph convolution with 1D convolution. This unique combination allows the model to effectively capture correlations between spatial and temporal dimensions, enabling a comprehensive understanding of the interplay between space and time in the data.

\item \textbf{GWNET} \cite{gwnet}: GWNET creates an adaptive adjacency matrix to capture spatial correlations and uses 1D dilated causal convolution to capture temporal dependence.

\item \textbf{ASTGCN} \cite{guo2019attention}: ASTGCN leverages the power of attention-based mechanisms and a spatio-temporal convolution system to dynamically capture spatio-temporal correlations within the data. By incorporating attention, the model can focus on relevant information and effectively model various temporal properties of traffic flows. 

\item \textbf{MTGNN} \cite{mtgnn}: MTGNN adopts a message-passing framework to effectively model the temporal dynamics of graph-structured data. It achieves this by aggregating information from spatially neighboring nodes and past time steps. By leveraging this approach, MTGNN captures the interdependencies and changes over time, enabling a comprehensive understanding of the data's temporal dynamics.

\item \textbf{AGCRN} \cite{bai2020adaptive}: AGCRN incorporates two key modules, namely Node Adaptive Parameter Learning and Data Adaptive Graph Generation, to automatically infer inter-dependencies in traffic series and capture node-specific patterns.


\item \textbf{STGNCDE} \cite{stgncde}: STGNCDE is an innovative spatio-temporal graph neural controlled differential equation model that leverages two neural control differential equations to process both spatial and sequential data. 

\end{itemize}

\textbf{Metrics.} To assess the performance of the above-mentioned models in spatio-temporal graph forecasting on our dataset, we employ the metrics of Mean Absolute Error (MAE) and Root Mean Squared Error (RMSE).

\section{Datasheet of Dataset}
\subsection{Motivation}

\mylist{\textbf{For what purpose was the dataset created?} Was there a specific task in mind? Was there a specific gap that needed to be filled? Please provide a description.}

\par To meet the rising calling for datasets in the field of last-mile delivery research, we propose \texttt{LaDe}, the first industry-scale multipurpose real-world dataset. Compared with existing public datasets, \texttt{LaDe} has serval merits: (1) large-scale, it consists of millions of packages, which can serve as a data foundation for learning-based algorithms in last-mile delivery. (2) Comprehensive information, the dataset contains more comprehensive features, which enables the data to support multiple research tasks. (3) Scenario diversity, it contains the data from both the package pick-up and delivery scenarios. Researchers can use the two sub-datasets to study the different work patterns of couriers in different scenarios.

\mylist {\textbf{Who created the dataset (e.g., which team, research group) and on behalf of which entity (e.g., company, institution, organization)?}}

\par The dataset was created by Artificial Intelligence Department, Cainiao Network.

\mylist { \textbf{Who funded the creation of the dataset?} If there is an associated grant, please provide the name of the grantor and the grant name and number.}

\par No.

\subsection{Composition}
\mylist { \textbf{What do the instances that comprise the dataset represent (e.g., documents, photos, people, countries)?} Are there multiple types of instances (e.g., movies, users, and ratings; people and interactions between them; nodes and edges)? Please provide a description.}

\par The instances are packages picked up/delivered in the last-mile delivery.

\mylist {\textbf{How many instances are there in total (of each type, if appropriate)?}}

\par There are {10,667k} instances in \texttt{LaDe}, where an instance represents the features of a package.

\mylist { \textbf{Does the dataset contain all possible instances or is it a sample (not necessarily random) of instances from a larger set?} If the dataset is a sample, then what is the larger set? Is the sample representative of the larger set (e.g., geographic coverage)? If so, please describe how this representativeness was validated/verified. If it is not representative of the larger set, please describe why not (e.g., to cover a more diverse range of instances, because instances were withheld or unavailable).}

\par The dataset is a sample of instances. We first randomly select serval regions in a city, then collect all the packages in that region within a certain period. Note that for each region, the dataset contains all possible instances within the given time period. To further increase the diversity of the dataset, five cities with different populations are selected and recorded.

\mylist { \textbf{What data does each instance consist of?} “Raw” data (e.g., unprocessed text or images) or features? In either case, please provide a description.}

\par The format of each instance in \texttt{LaDe-P} is \textit{(package\_id, time\_window\_start, time\_window\_end, lng, lat, city, aoi\_id, aoi\_type, courier\_id, accept\_time, accept\_gps\_time, accept\_gps\_lng, accept\_gps\_lat, pickup\_time, pickup\_gps\_time, pickup\_gps\_lng, pickup\_gps\_lat, ds)}.  

\par The format of each instance in \texttt{LaDe-D} is \textit{(package\_id,  lng, lat, city, aoi\_id, aoi\_type, courier\_id, accept\_time, accept\_gps\_time, accept\_gps\_lng, accept\_gps\_lat, delivery\_time, delivery\_gps\_time, delivery\_gps\_lng, delivery\_gps\_lat, ds)}. 

\par For the detailed description of each field, please refer to Table~\ref{tab:pickup_data_field} and Table~\ref{tab:delivery_data_field} in Appendix~\ref{appendix:data_detail}.

\mylist{ \textbf{Is there a label or target associated with each instance?} If so, please provide a description.}
\par Since the dataset is proposed to support multiple tasks in last-mile delivery, for easy use and flexibility, a label for a specific task is not contained in one instance. However, it is easy to construct the label for different research purposes from the raw information.  Take the estimated time of arrival prediction as an example. The actual arrival time (in this case, the label) can be calculated by the difference between the got\_time and query\_time.

\mylist{ \textbf{Is any information missing from individual instances?}  If so, please provide a description, explaining why this information is missing (e.g., because it was unavailable). This does not include intentionally removed information, but might include, e.g., redacted text.}
\par Some instances lack the courier's location when accepting/finishing the package, i.e.,  accept\_gps\_lng, accept\_gps\_lat. The corresponding information is massing in the real system.

\mylist{ \textbf{Are there recommended data splits (e.g., training, development/validation, testing)?} If so, please provide a description of these splits, explaining the rationale behind them.}
\par For all the tasks conducted in the paper (i.e., route prediction, time prediction, and spatio-temporal graph forecasting), we split the data into 6:2:2 according to the time as the training set, validation set, and test set.

\mylist{ \textbf{Are there any errors, sources of noise, or redundancies in the dataset?} If so, please provide a description.}
\par No.

\mylist{  \textbf{Is the dataset self-contained, or does it link to or otherwise rely on external resources (e.g., websites, tweets, other datasets)?} If it links to or relies on external resources, a) are there guarantees that they will exist, and remain constant, over time; b) are there official archival versions of the complete dataset (i.e., including the external resources as they existed at the time the dataset was created); c) are there any restrictions (e.g., licenses, fees) associated with any of the external resources
that might apply to a dataset consumer? Please provide descriptions of all external resources and any restrictions associated with them, as well as links or other access points, as appropriate.}
\par The dataset is entirely self-contained.

\mylist{ \textbf{Does the dataset contain data that might be considered confidential (e.g., data that is protected by legal privilege or by doctor–patient confidentiality, data that includes the content of individuals’ nonpublic communications)?} If so, please provide a description.}
\par No.

\mylist{ \textbf{Does the dataset contain data that, if viewed directly, might be offensive, insulting, threatening, or might otherwise cause anxiety?} If so, please describe why.}
\par No.

\par \textbf{Does the dataset identify any subpopulations (e.g., by age, gender)?} If so, please describe how these subpopulations are identified and provide a description of their respective distributions within the dataset.
\par No.

\par \textbf{Is it possible to identify individuals (i.e., one or more natural persons), either directly or indirectly (i.e., in combination with other data) from the dataset?} If so, please describe how.
\par Our data has been strictly desensitized and cannot be linked to real individuals.

\mylist{ \textbf{Does the dataset contain data that might be considered sensitive in any way (e.g., data that reveals race or ethnic origins, sexual orientations, religious beliefs, political opinions or union memberships, or locations; financial or health data; biometric or genetic data; forms of government identification, such as social security numbers; criminal history)?} If so, please provide a description.}
\par No.

\subsection{Collection Process}
\mylist{\textbf{How was the data associated with each instance acquired? Was the data directly observable (e.g., raw text, movie ratings), reported by subjects (e.g., survey responses), or indirectly inferred/derived from other data (e.g., part-of-speech tags, model-based guesses for age or language)?} If the data was reported by subjects or indirectly inferred/derived from other data, was the data validated/verified? If so, please describe how.}
\par The data was observable from the courier's pick-up/delivery data on the Cainiao platform.

\mylist{ \textbf{What mechanisms or procedures were used to collect the data (e.g., hardware apparatuses or sensors, manual human curation, software programs, software APIs)?} How were these mechanisms or procedures validated?}
\par The data is collected by the software program in the Cainiao platform.

\mylist{ \textbf{If the dataset is a sample from a larger set, what was the sampling strategy (e.g., deterministic, probabilistic with specific sampling probabilities)?}}
\par We pick out serval cities and randomly select regions in different cities.

\mylist{ \textbf{Who was involved in the data collection process (e.g., students, crowd workers, contractors) and how were they compensated (e.g., how much were crowdworkers paid)?}}
\par The employees in Cainiao.

\mylist{ \textbf{Over what timeframe was the data collected?} Does this timeframe match the creation timeframe of the data associated with the instances (e.g., recent crawl of old news articles)? If not, please describe the timeframe in which the data associated with the instances was created.}
\par This data was extracted from the Cainiao platform between May and November of a recent year.

\mylist{ \textbf{Did you collect the data from the individuals in question directly, or obtain it via third parties or other sources (e.g., websites)?}}
\par The data was collected from the Cainiao platform.

\subsection{Preprocessing/cleaning/labeling}
\mylist{\textbf{Was any preprocessing/cleaning/labeling of the data done (e.g., discretization or bucketing, tokenization, part-of-speech tagging, SIFT feature extraction, removal of instances, processing of missing values)?} If so, please provide a description. If not, you may skip the remaining questions in this section.}
\par We anonymize the courier's ID and package ID to protect user privacy. And we applied perturbations to the latitude and longitude points collected in the data.  The accuracy of the latitude and longitude is limited to 10 meters. 

\mylist{\textbf{Was the “raw” data saved in addition to the preprocessed/cleaned/labeled data (e.g., to support unanticipated future uses)?} If so, please provide a link or other access point to the “raw” data.}
\par No.

\subsection{Uses}
\mylist{ \textbf{Has the dataset been used for any tasks already?} If so, please provide a description.}
\par No.

\mylist{ \textbf{Is there a repository that links to any or all papers or systems that use the dataset?} If so, please provide a link or other access point.}
\par Yes.

\mylist{ \textbf{What (other) tasks could the dataset be used for?}}
\par The dataset can be used for route prediction, estimated time of arrival prediction, spatio-temporal graph forcasting, and route optimization. See section~\ref{sec:disscuss_of_other_tasks} for more details.


\subsection{Distribution}

\par \mylist{\textbf{How will the dataset will be distributed} (e.g., tarball on website, API, GitHub)? Does the dataset have a digital object identifier (DOI)?}
\par The dataset will be made available on the internet.  There will be a corresponding Hugging Face repository associated with the dataset, and code on how to use the dataset and baseline methods.

\mylist{ \textbf{When will the dataset be distributed?}}

\par The dataset is available to the reviewers and the public along with the submission with a companion Hugging Face repository.

\mylist{ \textbf{Will the dataset be distributed under a copyright or other intellectual property (IP) license, and/or under applicable terms of use (ToU)?} If so, please describe this license and/or ToU, and provide a link or other access point to, or otherwise reproduce, any relevant licensing terms or ToU, as well as any fees associated with these restrictions.}

\par This dataset is licensed under a CC BY-NC 4.0 International License \footnote{https://creativecommons.org/licenses/by-nc/4.0/}. There is a request to cite the corresponding paper if the dataset is used.

\par \textbf{Have any third parties imposed IP-based or other restrictions on the data associated with
the instances?} If so, please describe these restrictions, and provide a link or other access point
to, or otherwise reproduce, any relevant licensing terms, as well as any fees associated with these
restrictions.
\par No.

\par \textbf{Do any export controls or other regulatory restrictions apply to the dataset or to individual
instances?} If so, please describe these restrictions, and provide a link or other access point to, or
otherwise reproduce, any supporting documentation.
\par No.

\subsection{Maintenance}
\mylist{ \textbf{Who will be supporting/hosting/maintaining the dataset?}}
\par The employee in Cainiao will host the dataset on Hugging Face.

\mylist{ \textbf{How can the owner/curator/manager of the dataset be contacted (e.g., email address)?}}
\par The authors can be contacted via their emails mentioned in the paper.

\mylist{ \textbf{Is there an erratum?} If so, please provide a link or other access point.}
\par Not to our best knowledge.

\mylist{ \textbf{Will the dataset be updated (e.g., to correct labeling errors, add new instances, delete instances)?} If so, please describe how often, by whom, and how updates will be communicated to dataset consumers (e.g., mailing list, GitHub)?}
\par The corresponding Hugging Face page will be updated regularly.

\mylist{ \textbf{Will older versions of the dataset continue to be supported/hosted/maintained?} If so, please describe how. If not, please describe how its obsolescence will be communicated to dataset consumers.}
\par The old versions of the dataset will not be maintained. If we update the version of the dataset, we will put the specific details of the dataset update on the relevant Hugging Face.

\mylist{ \textbf{If others want to extend/augment/build on/contribute to the dataset, is there a mechanism for them to do so?} If so, please provide a description. Will these contributions be validated/verified? If
so, please describe how. If not, why not? Is there a process for communicating/distributing these
contributions to dataset consumers? If so, please provide a description.}

\par If others want to extend/augment/build on/contribute to the dataset, please contact the original authors about incorporating fixes/extensions.

\end{document}